# A Serious Game Approach for the Electro-Mobility Sector*

Bartolomeo Silvestri, Alessandro Rinaldi, Antonella Berardi, Michele Roccotelli, Simone Acquaviva
and Maria Pia Fanti, *Fellow, IEEE*

*Abstract*— Serious Games (SGs) represent a new approach to improve learning processes more effectively and economically than traditional methods. This paper aims to present a SG approach for the electro-mobility context, in order to encourage the use of electric light vehicles. The design of the SG is based on the typical elements of the classic "game" with a real gameplay with different purposes. In this work, the proposed SG aims to raise awareness on environmental issues caused by mobility and actively involve users, on improving livability in the city and on real savings using alternative means to traditional vehicles. The objective of the designed tool is to propose elements of fun and entertainment for tourists or users of electric vehicles in the cities, while giving useful information about the benefits of using such vehicles, discovering touristic and interesting places in the city to discover. In this way, the user is stimulated to explore the artistic and historical aspects of the city through an effective learning process: he/she is encouraged to search the origins and the peculiarities of the monuments.

A case study in the city of Bari, Italy, shows the application of the proposed SG tool.

## I. INTRODUCTION

Mobility management in the urban area is one of the main topics in the smart cities context. Transport sector is one of the main responsible for the production of greenhouse gases emissions. Several externalities are produced in this sector, such as air pollution (e.g., increase in particulate matter, nitrogen oxides), noise pollution, urban traffic congestion, accidents and parking shortages.

Smart mobility is a concept based on new solutions and innovations in order to improve the mobility citizens' behavior and at the same time the environmental condition in cities [1]-[8]. In this context, a smart mobility plan within the context of smart cities is a tacking challenge. Recent studies demonstrate as in Europe the most frequent trips are made by cars: it is about 60% of total transport modes used [9]. Some of the problems described are very dangerous for human health [10] and produce economic and social losses [11].

*This work is a part of the ELVITEN project. ELVITEN has received funding from the European Union's Horizon 2020 research & innovation programme under Grant Agreement no 769926. Content reflects only the authors' view and European Commission is not responsible for any use that may be made of the information it contains.

Bartolomeo Silvestri is with the Department of Mechanics, Mathematics and Management, Polytechnic University of Bari, 70125 Bari, Italy, (e-mail: bartolomeo.silvestri@poliba.it).

Alessandro Rinaldi, Antonella Berardi, Michele Roccotelli and Maria Pia Fanti are with the Department of Electrical and Information Engineering, Polytechnic University of Bari, 70125 Bari, Italy, (e-mail: alessandro.rinaldi,antonella.berardi,michele.roccotelli,mariapia.fanti@poliba.it).

Simone Acquaviva is with Polytechnic University of Bari, 70125 Bari, Italy, (e-mail: s.acquaviva@studenti.poliba.it).

In addition, public institutions and organizations are working to implement policy guidelines to support sustainability, such as the United Nations Agenda 2030 action plan that aims at preserving the planet and promote prosperity [12]. The European Commission objective of decarbonization in the Roadmap 2050 is to decrease of 80% the $CO_2$ emission since 1990: it focuses on reducing transport externalities by promoting public transport and banning the use of conventional internal combustion engine cars in cities by 2050. Therefore, the introduction of low environmental impact transport mode such as Electric Vehicles (EVs), the use of smaller, lighter and more specialized road passenger vehicles, such as Electric L-category Vehicles (EL-Vs) and of new strategies, such as vehicles sharing, combined with ICT tools, are the innovative solutions proposed in the smart mobility framework.

A significant percentage of urban travels is made by using light vehicles (L-Vs), which are smaller than cars and can ensure reduced travel time and less fuel consumption, as well as less time spent looking for parking. For these reasons, the introduction of EL-Vs in the smart city represents an important step towards a more sustainable urban mobility, in addition to reducing emissions and noises. The use of electric vehicles is still hampered by several factors, such as the penetration in the market, purchase cost, limited direct experience of people especially related to range anxiety and performance.

Therefore, it is very important to propose appropriate usage schemes, support services and ICT tools designed to facilitate and increase the use of such transport means.

Recent studies show how it is possible to modify the personal mobility behavior using incentive strategies [13] and gaming strategies [14]. The use of Gamification is very powerful in different fields, such as education, rehabilitation, health, military simulation, environmental sustainability, and others.

Serious games are defined as "the use of game design elements in non-game context" [15]. Also in the mobility sector, serious games have been developed to encourage the voluntary behavioral changes [16] towards sustainable solutions. In addition to mere entertainment the serious games have different purposes, such as teaching, sensitization and knowledge.

Moreover, SGs definitely represent a new approach for learning processes and they are more effective and cheaper than traditional information means, especially in the international business environment.

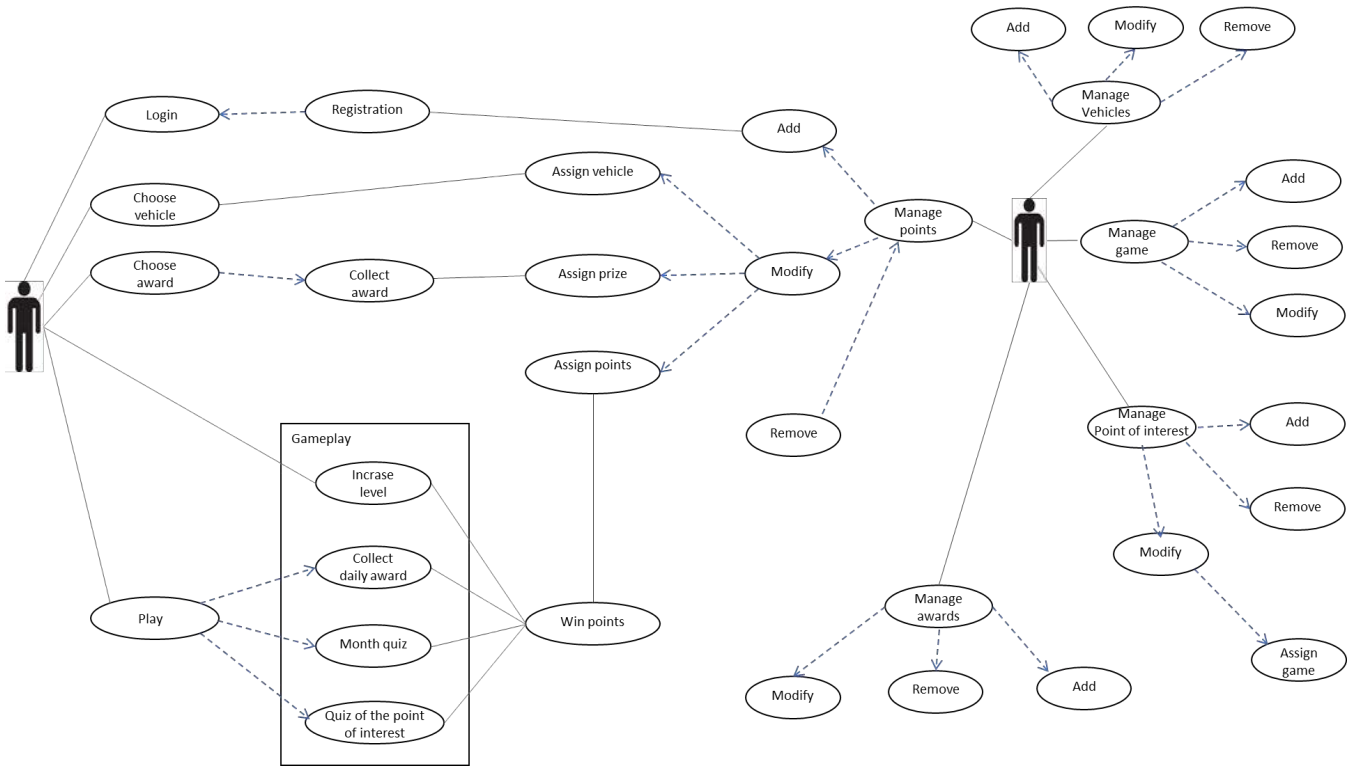

Figure 1. UML diagram of use cases

Nowadays, the use of ICT tools and smartphones [17] allows developing serious games more effectively. Several contributions are present in the literature about the use of serious games in the context of sustainable mobility. A serious game developed with an app to foster the use of sustainable transport mode as an alternative to private car, based on regular itineraries is proposed in [18]. The use of this approach is suggested by [19] to increase safe and sustainable mobility, and by [20] to improve transport maintenance and operations. Another approach is implemented in [21] with the use of gamification in a flexible transportation service for on-demand public transport.

In this paper a novel SG approach is proposed to promote electro-mobility. An incentive mechanism could be linked to increase user awareness of the environmental benefits deriving from electro-mobility, to enhance voluntary behavioral changes and sensitize towards historical-artistic places in the cities. The proposed SG approach aims at engaging users, by providing information and curiosities about the city and the use of green transport means, and at the same time encouraging to travel with EL-Vs in a more familiar way. In particular, a general and highly scalable methodology to define the ICT application requirement and architecture is presented, that can be also applied in different contexts. A preliminary version of the application has been developed and tested in the city of Bari in the context of the European project ELVITEN, that focuses on the analysis and demonstration of EL-Vs use in six pilot tests.

The rest of the paper is organized as follows: in Section II we present the methodology and the aims; the architecture of the serious games is presented in Section III; in Section IV we show the rules and functionalities of the app. In section V a case study showing a real test of the proposed app is described. Finally, Section VI reports the conclusions.

## II. SERIOUS GAME APPLICATION GOALS AND REQUIREMENTS

In this work, the proposed SG aims to raise user awareness regarding the environmental issues, such as $CO_2$ emissions and air pollution. Moreover, an important objective is pushing the citizens to improve the livability of the city by using alternative transportation means, such as public vehicles or light electric vehicles. Another desired educational goal is to introduce the historical-artistic side of the city, stimulating the curiosity of users by offering them information about monuments and historical places that very often remain unknown to the citizen or visiting tourist.

The objective of the designed tool is to have elements of fun and entertainment for tourists or users of EL-Vs in the cities, while giving useful information about the benefits of using EL-Vs, to discover touristic or interesting places in the city.

As common approach used in literature [22] - [27], we define the functional requirements on the basis of the "use cases" and their representation through the Unified Modeling Language (UML) [28]. In particular, use cases are adopted to provide a high-level or conceptual scenarios for deriving the requirements. This approach allows to identify the process flow, the interactions between a system and one or more actors, and "main" and "alternative flows" of steps/actions, providing a formal specification of the expected requirements. The UML Use Case diagram typically represents actors and use cases, and their interrelations [29].

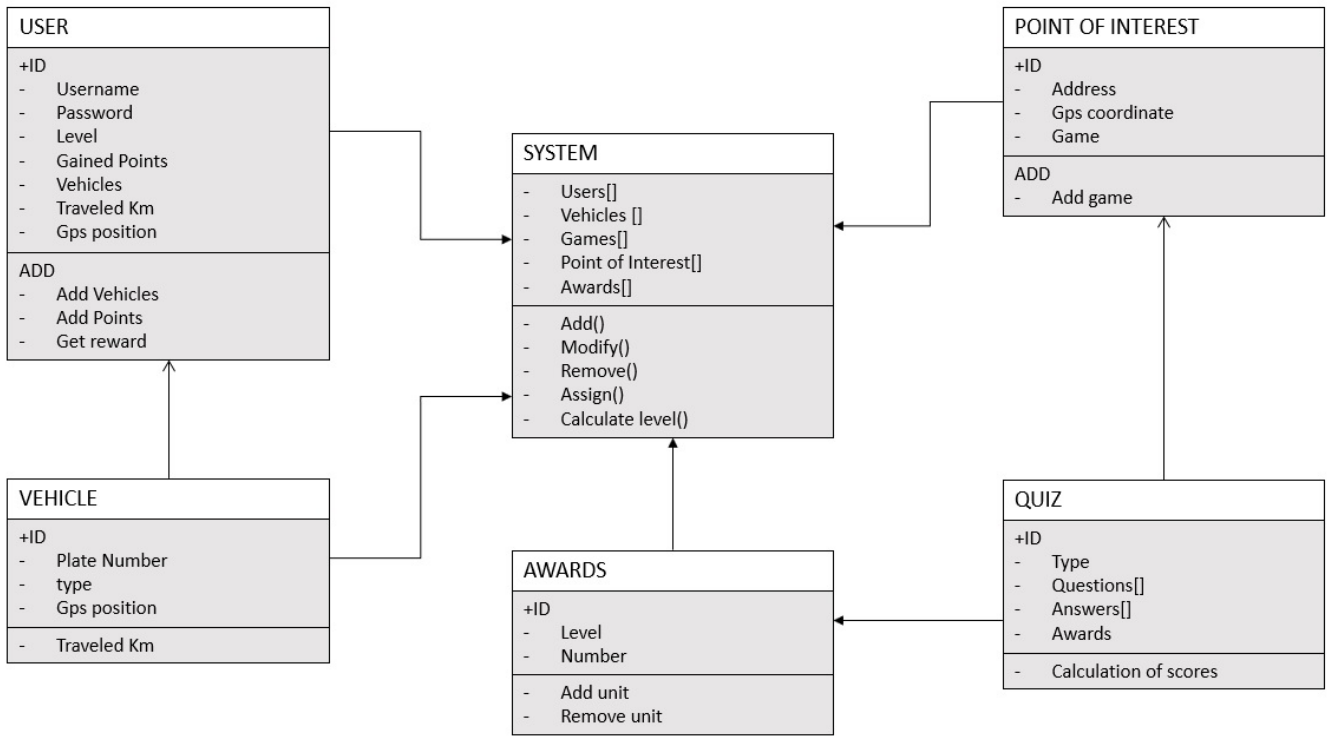

Figure 2. UML diagram of classes.

The UML use case diagram in Fig. 1 shows how people expect to use the app, by describing the main actors, the functionalities they require from the system and the relations among them. In particular, there are two actors: a) the user who plays with the application, (he/she can define the means of transport, play with the various proposals of the app and choose among the possible prizes); b) the system which manages and updates the games and the prizes to be offered to the user, and administers the users and their data such as associated transport vehicles and virtual wallet.

In addition, Fig. 2 shows the UML class diagram representing the main classes of the designed application.

More in detail, five classes define the SG architecture:

- the data of EL-V user and the related trip;
- the characteristics of the EL-Vs;
- the awards that can be obtained by using the app;
- the points of interest that are the main historical and cultural locations in the city;
- the questionnaires to be answered related to the previous points of interest.

To identify the use cases, the following aspects are analyzed:

- functionalities to be included;
- relations between the involved actors;
- identification of the users of each system;
- identification of the dependencies between the involved actors;

The defined use cases to be implemented through the application are the following:

- the user registration/authentication;
- the geolocation;
- the identification of points of interest;
- the access to the game and saving of the game play results.

The SG is based on two different processes: one linked to the user's GPS coordinates and one related to a continuous offline data flow.

III. ARCHITECTURE OF THE SERIOUS GAME APPLICATION

This section presents the adopted methodology to define, design and implement the architecture of the SG application. The used development environment is Android Studio, an integrated development environment (IDE) for the development of the Android platform. The chosen programming language is Java, an object-oriented programming language designed specifically to be as independent as possible from the execution platform. The application consists of a total of about 20 activities, which are the core components of an Android application. The way in which the activities interact with each other constitutes the application's working model. Each Activity consists of a Java file and an XML (eXtensible Markup Language) file.

The Manifest Android file is implemented for the structure of the app and includes some information such as: the permissions requested from the user, the Activity used, app theme and customization. The App contents are completely customized by using XML files. Then four XML files will be used by the App:

- LocationList.xml contains the points of interest defined in the demo site which will trigger quizzes. The number of places of interest, the related questions and the points (general or topic-related) for each correct answer are different for each city. Also, each achievement reached may lead to incentive points for the user.

- Geolocation.xml includes the following parameters: Id, name, GPS position, distance at which quiz is triggered, Id of the message shown when triggered.

- GameSettings.xml: contains the main indicators for local customization: languages, topics of the quizzes, achievements foreseen.

- MessagesList.xml: contains all the quizzes and messages that can be shown to the user during the game.

The geolocation function of the mobile device allows to activate the questionnaires when the user is near the monuments of the city included in the game. The main purpose of the data collection by the app is to have a clear vision of how much knowledge and experience of the electric vehicles and electro-mobility in general the citizens have. In particular, the scores obtained by the questionnaire answers will be linked to the incentives and awards. These data will be used to understand the awareness level of the citizens on the sustainable mobility with particular reference to the electro-mobility sector and how much the SG helps to increase the willingness to use EVs and EL-Vs for the future.

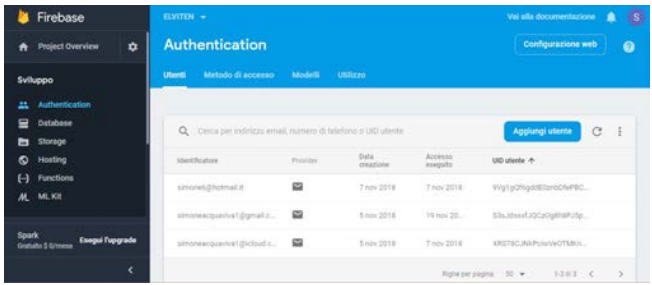

Figure 3. The Authentication page.

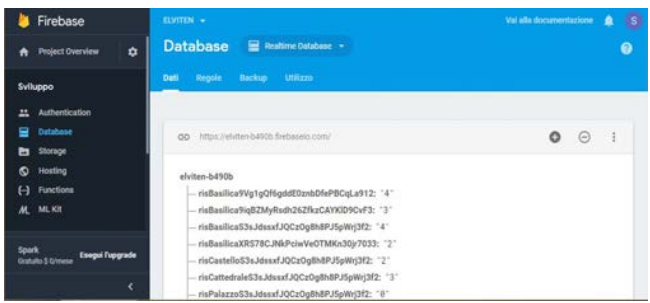

Figure 4. The database layout.

The platform used for authentication and data storage (Fig.3 and 4, respectively) is Google's Firebase: it is a platform that integrates a set of features for mobile apps on iOs and Android. These features can be divided into 4 groups: analytics, development, growth and monetization. In the Authentication section of the platform, it is possible to view registered users, identified by the user ID, as well as being able to check when these users have accessed the platform.

IV. SERIOUS GAME RULES AND FUNCTIONALITIES

The main functions implemented in the designed SG are the following:

1. The user has the possibility to create his own account, which is identified by e-mail, username and password. The login and registration functions are implemented within the application itself and the data are saved online on the Google Firebase service. Invalid registration/login values are communicated to the user by a Toast message. The user who logs in will remain logged in automatically: there is no need to authenticate each time the application is opened.

2. In a dedicated activity, the application captures the user's position and tracks it in real time. The position is marked on a map by a red marker that identifies and displays the address in which the user is located.

3. Places of interest are marked on the map, each one with a different colored marker. By clicking on one of them the user can connect directly to the Google Maps application and see the shortest route to reach the desired attraction.

4. When the user is less than 200m from one of these points, the application automatically detects it and starts the corresponding multiple-choice questionnaire (Fig. 5). This method requires three parameters: 1) starting point (latitude and longitude); 2) end point (latitude and longitude) and finally a float vector.

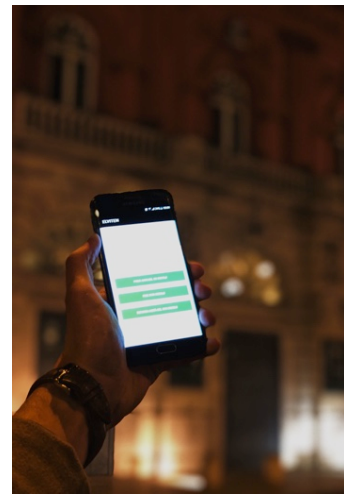

Figure 5. The questionnaire is activated.

5. At the end of the questionnaire the application shows a customized message that depends on the number of correct answers selected.

The SG allows saving the results in the Firebase database. Each response saved on the database can be changed only if the same user repeats the questionnaire and decides to modify the previous result.

6. On the main page of the application there is a link to the "Your results" activity. Here the user reads the saved score for each questionnaire. If the questionnaire has never been completed, the score field will be empty. The results of the questionnaires are visible in the Database section. Automatically, when a user decides to save a result, a unique string is created, defined by the name of the questionnaire that was linked to the user ID, which is unique.

## V. CASE STUDY: APPLICATION IN ELECTRO-MOBILITY CONTEXT

The proposed SG is applied to the electro-mobility context in the city of Bari, the capital city of the Apulia Region, Italy. This tool is designed to encourage the EL-V usage and promote the consequent discovery of places of interest in Bari. It can be considered as an alternative method to the traditional tourist guide.

The basic game idea is a mix of "Treasure Hunt" and "Trivial Pursuit" game concepts. A set of quizzes are proposed (with possible multimedia elements) which will be location-based, meaning that they will be triggered when the user is near to some points of interest in the city. The proposed questions are linked to specific places in the city (e.g.: a street, a monument, etc.) and about general information on EL-Vs. When a point of interest is reached, one or more quizzes will be triggered (i.e. will be proposed to the users).

In this way, the users can explore the city in order to discover the important places in a sort of treasure hunt that may help to discover "hidden gems" and beautiful places in the city, by using their electric light vehicles. Each quiz answered correctly leads to gain incentive scores related to the quiz topic.

After the login and authentication, the user can see on the map the places of interest, the available parking spaces and his/her current position (Fig. 6). Therefore, the SF motivates the user to move by using the EL-Vs. Consequently, the user familiarity with this type of vehicle increases. The questionnaires are customized for a specific target group of users in order to involve more actors and create interesting and pertinent questions.

To promote a gradual habit change, it is necessary to proceed step by step. For this reason, two different levels of game are proposed (easy and hard game level), to solve the problem of abandoning the game. Multiple questions proposals have been dedicated to each points of interest, with two/three possible answers for each question. Many questions are also thought to make the user experience more immersive. In particular, three question topics are identified: history, arts/show/trivia and EL-Vs (Fig. 7).

At the end of each quiz session, a screen appears with the indications of the collected scores (Fig. 8). The user's ability to save the results in his account on one hand allows him to accumulate scores (the collected scores allow to earn rewards).

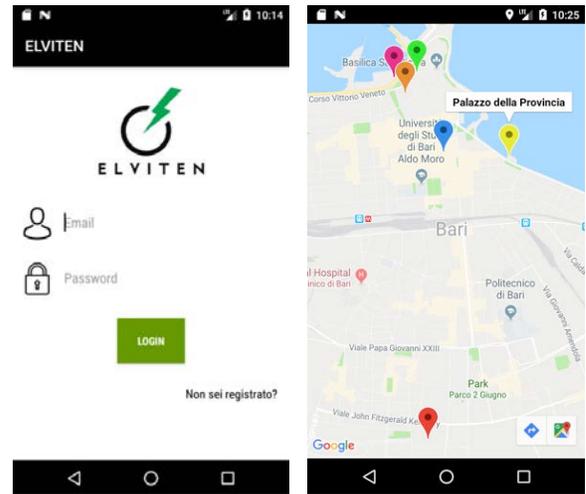

Figure 6. Screenshots of the SG.

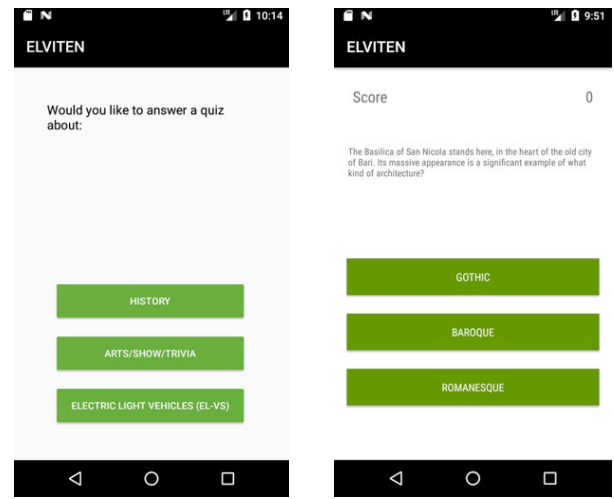

Figure 7. The question topics and questions pages.

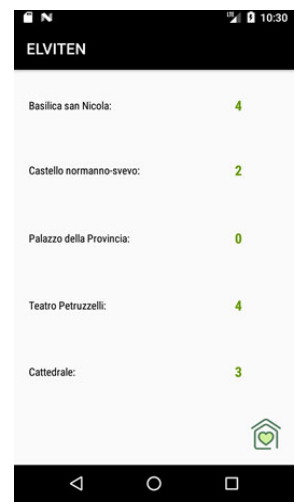

Figure 8. The score screen.

## VI. Conclusion

This paper proposes an innovative methodology for involving people in a systematic use of electro-mobility by employing SG. To this aim, an app is developed to involve tourists and users in EL-Vs usage, and at the same time, entertain and inform them about the benefits of electro-mobility and tourist attractions.

The user is stimulated to explore the artistic and historical aspects of the city through an effective learning process; he/she is encouraged to search the origins and the peculiarities of the monuments.

The app has all the most important features of a SG: it is simple, intuitive and allows a useful interaction with the users. Moreover, the SG pursues social functions such as stimulating the competition among the users and encouraging the knowledge of the artistic-historical side of the city. The user is stimulated, by the questions, in the search for the origins and the peculiarities of the monuments in order to gain additional scores.

In future works, a specific platform will be designed and linked with an incentive system to spend points earned through different rewards.

# NOTICE